\documentclass[conference]{IEEEtran}
\IEEEoverridecommandlockouts
\usepackage{cite}
\usepackage{amsmath,amssymb,amsfonts}
\usepackage{algorithmic}
\usepackage{graphicx}
\usepackage{textcomp}
\usepackage{xcolor}
\usepackage{booktabs} 
\usepackage{colortbl}
\usepackage{listings}
\usepackage{bytefield}
\usepackage{url}
\usepackage{pifont}

\newcommand\redding[1]{{\color{red}\large\ding{#1}}}
\newcommand{\One}{{\redding{202}}}
\newcommand{\Two}{{\redding{203}}}
\newcommand{\Three}{{\redding{204}}}
\newcommand{\Four}{{\redding{205}}}
\newcommand{\Five}{{\redding{206}}}
\newcommand{\Six}{{\redding{207}}}
\newcommand{\Seven}{{\redding{208}}}
\newcommand{\Eight}{{\redding{209}}}

\begin{document}

\title{Zest: REST over ZeroMQ}

\author{\IEEEauthorblockN{John Moore, Andr\'{e}s Arcia-Moret,\\
  Poonam Yadav, Richard Mortier}
\IEEEauthorblockA{\textit{University of Cambridge} \\
Cambridge, UK}
\and
\IEEEauthorblockN{Anthony Brown, Derek McAuley,\\
   Andy Crabtree, Chris Greenhalgh}
\IEEEauthorblockA{\textit{University of Nottingham} \\
Nottingham, UK}
\and
\IEEEauthorblockN{Hamed Haddadi}
\IEEEauthorblockA{\textit{Imperial College} \\
London, UK}
\and
\IEEEauthorblockN{Yousef Amar}
\IEEEauthorblockA{\textit{Queen Mary University} \\
London, UK}
}

\maketitle

\begin{abstract}
  In this paper, we introduce Zest (REST over ZeroMQ), a middleware technology
  in support of an Internet of Things (IoT). Our work is influenced by the
  Constrained Application Protocol (CoAP) but emphasises systems that can
  support fine-grained access control to both resources and audit information,
  and can provide features such as asynchronous communication patterns between
  nodes. We achieve this by using a hybrid approach that combines a RESTful
  architecture with a variant of a publisher/subscriber topology that has
  enhanced routing support. The primary motivation for Zest is to provide
  inter-component communications in the Databox, but it is applicable in other
  contexts where tight control needs to be maintained over permitted
  communication patterns.
\end{abstract}

\maketitle

\section{Introduction}
\label{s:introduction}

The goal behind Zest is to utilise middleware to improve on the features offered by the Constrained Application Protocol (CoAP)~\cite{coap-bormann,coap-kovatsch} by providing:
\begin{itemize}
\item Encryption as standard
\item Access control through Macaroons
\item Support for auditing communication across nodes
\item Support for asynchronous communication between nodes
\end{itemize}
We chose to build our solution using ZeroMQ\footnote{\url{http://zeromq.org/}} because of its flexibility to support different topologies such as brokerless communication and for its simple abstraction over traditional TCP sockets. Other reasons we adopted ZeroMQ included its support for secure connections based on elliptic-curve cryptography and that it is well supported across a variety of platforms and programming languages.
\begin{figure}
  \centering
  \includegraphics[width=\linewidth]{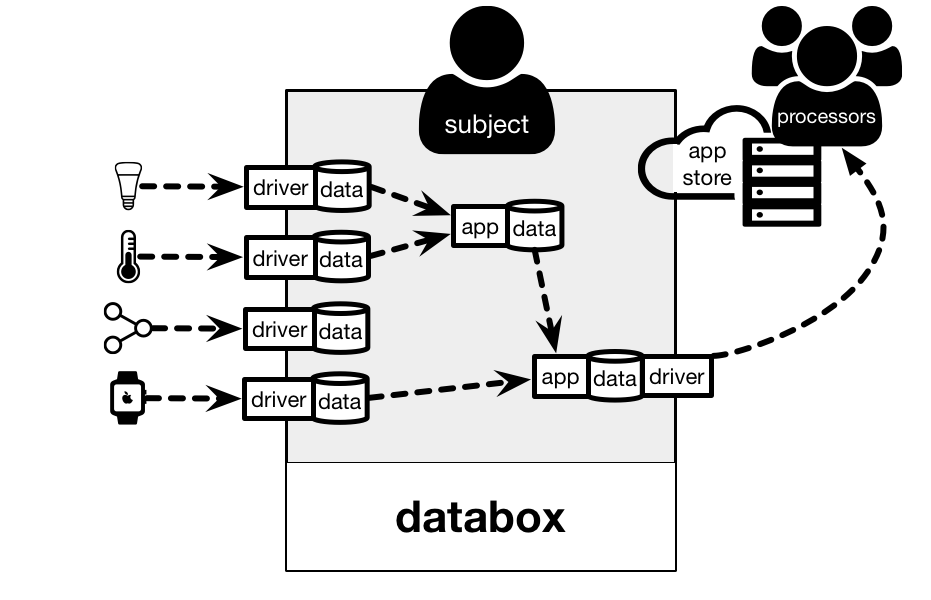}
  \caption{\label{f:databox}Databox~architecture.}
\end{figure}
Zest forms the core protocol within the Databox project~\cite{databox}, which we envisage being instantiated in the form-factor of a set-top box or similar. All components are encapsulated as Docker containers.\footnote{\url{https://docker.com/}} Databox hosts third-party computations as \emph{Apps}, while external devices such as sensors interface to the Databox~via \emph{Drivers} responsible for interacting with the external device through reads and writes to an associated \emph{store}, a light-weight time-series database.  Zest's requirements therefore are to support this highly controlled communication model, in a relatively resource-constrained environment, where operations must be logged for subsequent audit and where data transfers should be authenticated and protected in flight. 
The Databox communication model not only involves encryption across communication channels but also requires support for fine-grained access to resources. We therefore developed the Zest protocol to support these features as standard, including support for audit information to be pushed to any App permitted to receive it.  Data is isolated within Databox by enforcing that each Driver may write only to its own Store, and Apps must request permission on installation to be able to access a Store. If the user grants permission, the \emph{App} receives a set of access tokens (formatted as Macaroons~\cite{macaroons}) which it can subsequently present to the Store to verify its access is allowed. Data may only be communicated to a third-party service through an \emph{Export Driver}, subject to the user granting appropriate permission when installing the \emph{App}. Available Stores are registered in a HyperCat catalogue on installation, so they can be discovered by other Apps.\footnote{\url{https://hypercat.io/}} Each store provides a \textit{RESTful} API supporting JSON, text and binary data across the Zest protocol. Underlying storage is implemented using the Irmin~\cite{irmin} system using a \emph{git}-structured backend. This supports a key design goal of Databox, to provide accountability of data stored and accessed, by using the commit history of the git-based storage system to provide a detailed account of all mutations to data.  In the remainder of this paper we discuss related work~(\S\ref{s:relatedwork}), detail the Zest protocol~(\S\ref{s:protocol}) and architecture~(\S\ref{s:architecture}),  and conclude~(\S\ref{s:conclusion}).

\section{Related Work}
\label{s:relatedwork}

A number of alternative technologies exist which can be used to support an Internet of Things, however, the baseline of our work corresponds to the Constrained Application Protocol (CoAP) for which the need for integration with enterprise infrastructure has recently motivated its development over TCP~\cite{RFC8323,GOMEZ}. The goal of the CoAP standard was to bring a \textit{RESTful} experience to the Internet of Things. As such, its design targets low-end devices including micro-controllers and it runs over UDP to provide more light-weight communication. However, the decision to use UDP sometimes introduces complexity such as when operating over networks that use NAT or when required to handle large message payloads efficiently. Upgrading CoAP to TCP required changes to the protocol structure to, for example, to accommodate reliable communication. However, complexities remain. The CoAP standard supports the concept of observing a resource so that data can be pushed back to client nodes. This type of interaction is intended to be supported using WebSockets, but this requires supporting an additional protocol. 

The basic concept of supporting a RESTful architecture using ZeroMQ is not new.\footnote{\url{https://rfc.zeromq.org/spec:40/XRAP/}} Novelty in Zest arises from adoption of CoAP's approach plus additional features introduced in \S\ref{s:introduction}. A number of technologies exist that do not adopt a RESTful approach but are suitable for building and deploying IoT solutions. Some make assumptions on the topology that can be used between communicating nodes and some place restrictions on the data format used within messages. For example, MQTT\footnote{\url{http://mqtt.org/}} is a publisher/subscriber messaging protocol where clients communicate via a server known as a broker. A client can connect as either a publisher, subscriber or both publisher and subscriber. Brokers are topic based. A publisher will write data for a specific topic which any subscriber of that topic will receive. Topics can be subscribed to on a hierarchical basis with optional wildcard filtering within the topic path. Security can be implemented over TCP using SSL/TLS.

Protocol Buffers\footnote{\url{https://developers.google.com/protocol-buffers//}} was created by Google to be used for machine-2-machine communication between their servers. Its binary on-the-wire format provided a more light-weight alternative to serialising text-based formats such as XML. Google also developed Cap'n Proto\footnote{\url{https://capnproto.org/}} as an improvement over Protocol Buffers in terms of its performance and also included the addition of RPC capabilities. The serialisation approach of both Protocol Buffers and Cap'n Proto is similar to that of Abstract Syntax Notation One (ASN.1)~\cite{x208} where the protocol is specified in a platform independent language which needs to be parsed into target language code with library support that can generate a binary on-the-wire representation of the data. Binary protocols are efficient for machine-2-machine communication, especially when it comes to transferring numeric data, however, they enforce an on-the-wire format. Using RPC technologies you are not restricted to use a broker to mediate communications like MQTT but you still need to design suitable high-level modes of interaction that end-users or developers can understand. Choosing the correct technology depends very much on the use-case. For example, a traditional REST model is a good approach for database read and writes but is not well suited if a client needs to continually poll a resource to receive an update. Therefore, technologies such as CoAP and HTTP/2~\cite{RFC7540} help address these issues by describing support to push data back to clients. The Zest protocol takes a more hybrid approach and borrows from traditional pub/sub technologies such as MQTT to deploy a broker to handle such use-cases. The following section will describe the Zest protocol in more detail.

\section{Protocol}
\label{s:protocol}

Table~\ref{f:message} shows the structure of a Zest message. A message must consist of at least a header, however, the token, options and payload are all optional and depend on the type of message being sent.

\begin{table}
 \centering
  \begin{bytefield}[bitwidth=0.6em, bitheight=2.3em]{32}
    \bitheader{0,7,8,15,16,31} \\

    \begin{rightwordgroup}{Header}
      \bitbox{8}{Code} & \bitbox{8}{Option Count} & \bitbox{16}{Token Length \\ (network order)}
    \end{rightwordgroup} \\

    \begin{rightwordgroup}{Token \\ (optional)}
      \wordbox[tlr]{1}{token} \\
      \wordbox[blr]{1}{$\cdots$}
    \end{rightwordgroup} \\

    \begin{rightwordgroup}{Options \\ (optional)}
      \bitbox{16}{Option Code \textit{1} \\ (network order)} \bitbox{16}{Length \\ (network order)}\\

      \bitbox{32}{value}\\

	\skippedwords \\
     \wordbox[lrb]{1}{} \\

      \bitbox{16}{Option Code \textit{n} \\ (network order)} \bitbox{16}{Length \\ (network order)}\\
      \bitbox{32}{value}
    \end{rightwordgroup} \\

    \begin{rightwordgroup}{Payload \\ (optional)}
      \wordbox[tlr]{1}{payload} \\
      \wordbox[blr]{1}{$\cdots$}
    \end{rightwordgroup} \\

  \end{bytefield}
  \caption{\label{f:message}Protocol structure}
\end{table}

A header is made up of 32 bits. The first 8 bits of the header are used to set the type of message used being either a request message or a response message.
Table \ref{f:codes} lists the possible request and response codes.

\begin{table}
\centering
\begin{tabular}{|l|l|l|}
\hline
\rowcolor[HTML]{C0C0C0}
\textbf{Type} & \textbf{Code} & \textbf{Meaning} \\ \hline\hline
REQ $\rightarrow$ & 1             & GET              \\ \hline
REQ $\rightarrow$ & 2             & POST             \\ \hline
REQ $\rightarrow$ & 4             & DELETE           \\ \hline\hline

RESP $\leftarrow$  & 65            & Acknowledge (POST)                  \\ \hline
RESP $\leftarrow$ & 66            & Acknowledge (DELETE)                \\ \hline
RESP $\leftarrow$ & 69            & Acknowledge with payload (GET/POST) \\ \hline \hline

RESP $\leftarrow$ & 128           & Bad request                         \\ \hline
RESP $\leftarrow$ & 129           & Unauthorised                        \\ \hline
RESP $\leftarrow$ & 134           & Not acceptable                      \\ \hline
RESP $\leftarrow$ & 141           & Request entity too large            \\ \hline
RESP $\leftarrow$ & 143           & Unsupported content format          \\ \hline
RESP $\leftarrow$ & 160           & Internal server error               \\ \hline
RESP $\leftarrow$ & 163           & Service unavailable                 \\ \hline
\end{tabular}
\newline
\caption{Request and Response codes}
\label{f:codes}
\end{table}

The following 8 bits of the header specify how many options have been encoded.
Options are used to configure a message and provide extra information such as specify the content type of the payload if one exists.
The first 16 bits specify which option is being encoded based on its value shown in table~\ref{f:option_coding}.

\begin{table}
\centering
\begin{tabular}{|c|l|l|}
\hline
\rowcolor[HTML]{C0C0C0}
\textbf{code} & \textbf{meaning}        & \textbf{value/type}                        \\ \hline
3    & uri\_host       & string                                \\ \hline
6    & observe        & string set to 'data', 'audit' or 'notify'                             \\ \hline
11   & uri\_path       & string                                \\ \hline
12   & content\_format & unsigned integer based on type        \\ \hline
14   & max\_age        & unsigned integer representing seconds \\ \hline
2048   & public\_key        & string \\ \hline
\end{tabular}
\newline
\caption{Option encoding}
\label{f:option_coding}
\end{table}
Options are encoded in a tag, length, value sequence where the length refers to the number of bytes required to store the value.
The type of value encoded varies depending on the option. For example, to encode a \textit{content\_format} option requires encoding an unsigned integer representing the content type such that 0 represents text, 42 represents binary and 50 represents JSON.

\begin{table*}
\centering
  \begin{tabular}{l|c|c|c|c|c|c|}
    \cline{2-7}
    & \cellcolor[HTML]{C0C0C0}\textbf{GET REQ} & \cellcolor[HTML]{C0C0C0}\textbf{GET RESPONSE} & \cellcolor[HTML]{C0C0C0}\textbf{POST REQ} & \cellcolor[HTML]{C0C0C0}\textbf{POST RESP} & \cellcolor[HTML]{C0C0C0}\textbf{DELETE REQ} & \cellcolor[HTML]{C0C0C0}\textbf{DELETE RESP} \\ \hline
    \multicolumn{1}{|l|}{\cellcolor[HTML]{C0C0C0}\textit{uri\_path}}       & x                                        &                                               & x                                         &                                            & x                                           &                                              \\ \hline
    \multicolumn{1}{|l|}{\cellcolor[HTML]{C0C0C0}\textit{uri\_host}}       & x                                        &                                               & x                                         &                                            & x                                           &                                              \\ \hline
    \multicolumn{1}{|l|}{\cellcolor[HTML]{C0C0C0}\textit{content\_format}} & x                                        & x                                             & x                                         & +                                          & x                                           &                                              \\ \hline
    \multicolumn{1}{|l|}{\cellcolor[HTML]{C0C0C0}\textit{observe}}         & +                                        &                                               &                                           &                                            &                                             &                                              \\ \hline
    \multicolumn{1}{|l|}{\cellcolor[HTML]{C0C0C0}\textit{max\_age}}        & +                                        &                                               &                                           &                                            &                                             &                                              \\ \hline
    \multicolumn{1}{|l|}{\cellcolor[HTML]{C0C0C0}\textit{public\_key}}        &                                         & +                                              &                                           &                                            &                                             &                                              \\ \hline
  \end{tabular}
  \newline
  \caption{\label{f:options}Options per request or response: x is mandatory, + is optional}
\end{table*}

The Zest protocol supports POST, GET and DELETE to a specific endpoint specified through a path. As previously described, each method is configured by setting specific protocol options as shown in figure \ref{f:options}. The POST method is used to add data to a resource such as a database or provide data to update a dashboard. Unlike the CoAP standard, a POST method can not only acknowledge the status of the request but can also return a payload of data. A POST response message might contain an option specifying the content of its payload. A GET method is used to request data from a resource such as a database. The GET method has a variant to support observing a resource. This is used to obtain information suitable for logging and auditing interaction across a Zest deployment. Observing a resource sets up a connection to the router endpoint of a Zest node which remains active until a specific expiry period is reached as dictated by the \textit{max\_age} option. If the \textit{max\_age} option is not provided a default of 60 seconds is assumed. In addition, a \textit{max\_age} value of 0 indicates the connection should not expire. As with all method calls, their usage should be controlled through suitably minted access tokens. The response to a GET method always contains a payload so must have the \textit{content\_format} option set in the response. Finally, the DELETE method is used to remove data from a resource such as a database. Note that although the DELETE request will not contain any payload data it still requires the \textit{content\_format} option to be set to identify the type of data that will be deleted. A DELETE response has no options or payload so consists of a header only.

In addition to the standard REST protocol there is a meta-protocol which is delivered across the router endpoint of a Zest node. The meta-protocol is a simple character separated format delivered as the payload of a Zest acknowledgement to a notification or observation request (see \ref{s:architecture} for more details). Observations have 3 formats based on the request: data, audit, notify. Whereas a notification has a single format. For example, the result of observing data posted to path \url{/kv/foo/bar} might look like:
\begin{lstlisting}[basicstyle=\scriptsize]
#timestamp #uri-path #content-format #data
1521554211213 /kv/foo/bar json {"room": "lounge", "value": 1}
\end{lstlisting}
Note that the first line contains a comment to represent the structure and is not included in the protocol. Further details and examples of the different formats are available online.\footnote{\url{https://github.com/me-box/zestdb/tree/master/docs\#observation}}

\section{Architecture}
\label{s:architecture}

\begin{figure}
  \centering
  \includegraphics[width=\linewidth]{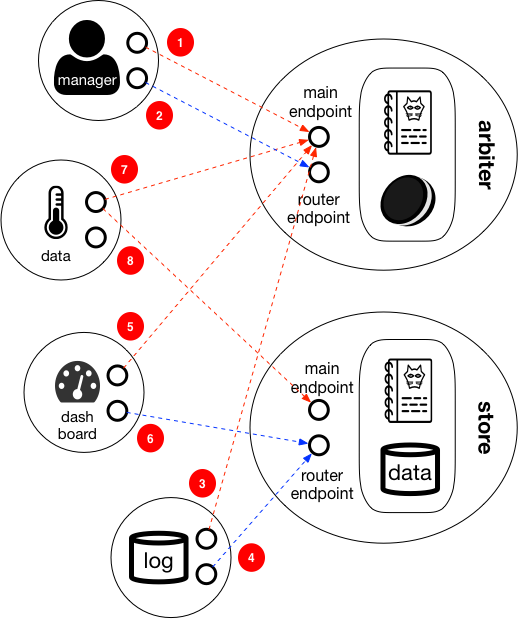}
  \caption{\label{f:generic_arch}Node interaction across a Zest deployment: \One~Manager connects to Arbiter to set up permissions for minting tokens, \Two~Manager subscribes to Arbiter to receive audit information, \Three~Logger connects to Arbiter to request audit logging token for Store, \Four~Logger subscribes to Store using token to receive audit information, \Five~Dashboard connects to Arbiter to request data logging token for Store, \Six~Dashboard subscribes to Store using token to receive copy of data posted, \Seven~Sensor connects to Arbiter to request token for posting data to Store, \Eight~Sensor posts data to Store using token}
\end{figure}

Figure \ref{f:generic_arch} depicts a simple Zest deployment where a number of client nodes interact with a server node (store) which has data storage and retrieval functionality. For example, ZestDB\footnote{\url{https://me-box.github.io/zestdb/}} is a storage node which provides both key/value and time-series functionality. Access to this node is controlled through tokens which need to be obtained from a privileged server node known as the arbiter. To bootstrap a Zest deployment requires a single privileged node depicted in Figure~\ref{f:generic_arch} as the manager node. The manager node is required to grant permissions on the arbiter node. This sets up which tokens can be minted and which nodes can request them. For example, at step~\Three~the log node connects to the arbiter to request a token which it will later present to the store. The architecture is flexible as it can be connected up in different ways. The sensor node in this deployment is not receiving any logging data but it could request this from either the arbiter or store or both provided it had the required access token. Access tokens are implemented as Macaroons~\cite{macaroons} which are similar to traditional bearer tokens but allow restrictions (caveats) to be added for delegation. A Zest deployment must support at least 3 caveats: the REST method required, the resource path to access and the target identity of the node which will receive the token. Each server node in a Zest deployment must also provide a HyperCat to describe its state. For example, a store will describe what data sources it currently contains, whereas an arbiter will describe which nodes have been granted permissions to request tokens.

\subsection*{Observations and Notifications}
\label{ss:observations}

\begin{figure}
  \centering
  \includegraphics[width=\linewidth]{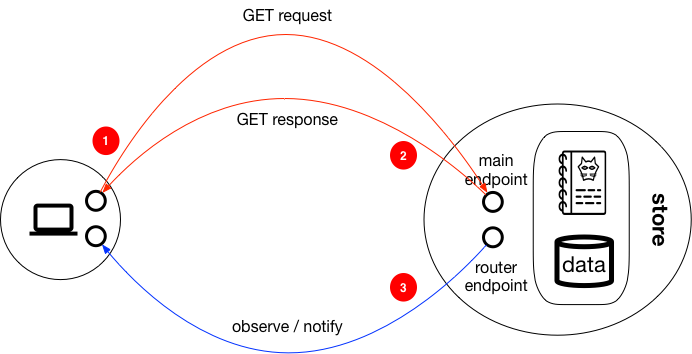}
  \caption{\label{f:observe}Sequence to set up an observation or notification.}
\end{figure}

Observations are used for logging/auditing data flowing through nodes, whereas notifications are used to support asynchronous communication between nodes. A Zest node provides two endpoints. The main endpoint uses a request/reply protocol to support the RESTful interface and is used to set up communication with the router endpoint, The router endpoint supports a router/dealer topology in ZeroMQ which is responsible for pushing data back to clients. A router/dealer topology in ZeroMQ differs from a traditional publisher/subscriber topology in that it maintains an internal routing table to ensure data is only routed to a single client connection rather than broadcast to multiple subscribers. Figure \ref{f:observe} summarises the sequence involved to set up an observation or notification. During an observation GET request the key of the server (e.g. store) is returned to the client node to allow secure communication to take place. In addition, this exchange also provides the client node with its unique identity in the form of a UUID which it is required to present when connecting to the router endpoint. A notification GET request follows a similar sequence, however, instead of obtaining a UUID from the server it generates its own identity in the form of a callback path for notification responses.
Notifications are used in conjunction with observations and support communication between two nodes interacting with a Zest store through a \textit{/notification/request} and \textit{/notification/response} path. Data is communicated through a store using a client/server paradigm with the store acting as a broker between the client and server nodes.
A client node can issue a request to a server node which can obtain the necessary information from an observation to respond back asynchronously with the result. Within Databox this allows an application developer to build an app or driver which uses the services of an existing app. For example, there may be an existing image processing app which specialises in offering face detection algorithms and this app has made this service available within the Databox system.
\begin{figure}
  \centering
  \includegraphics[width=\linewidth]{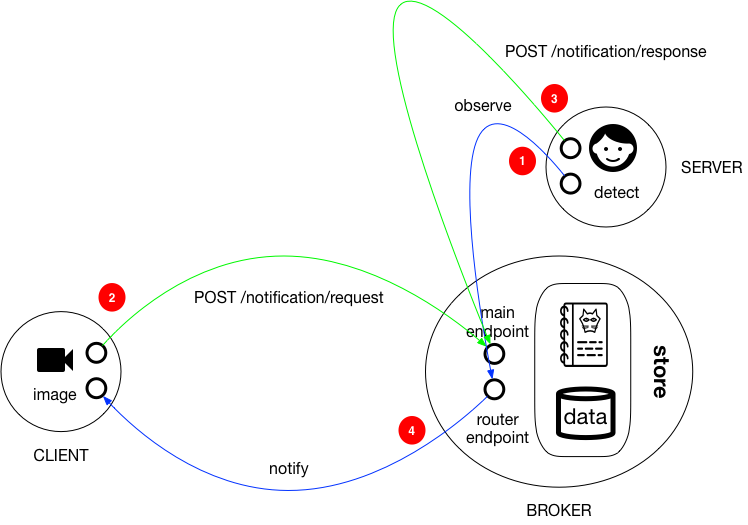}
  \caption{\label{f:notify}Asynchronous communication with a client node talking to server node through a store acting as a broker.}
\end{figure}
Figure \ref{f:notify} illustrates how a client capturing images is able to send them to a server node for facial recognition and receive back the results using the notification system.
The intermediary node (store) acts as a broker between client and server nodes to facilitate the communication.
This takes place over predefined paths on the broker which means both client and server can be controlled through access tokens and have no direct way of communicating with one another.
The sequence of steps involved could be summarised by the following interaction:
\begin{enumerate}
\item Server observes requests on broker path \\\textit{/notification/request/image\_capture/*}
\item Client POSTs image to broker path \\\textit{/notification/request/image\_capture/001}
\item Server carries out image processing and POSTs result to broker path \\\textit{/notification/response/image\_capture/001}
\item Client receives notification containing processed image through a previously established GET request to path \\\textit{/notification/response/image\_capture/001}
\end{enumerate}
Note, that steps 1 and 4 translate into the sequence previously described in figure \ref{f:observe}. In this example, the server node is using a wildcard to accept requests from any path with the \emph{/image\_capture/} prefix and therefore will accept the client request of \emph{/image\_capture/001}. The Zest protocol does not enforce a particular naming scheme, so a client is responsible for generating a unique request path to ensure that a unique callback is generated, for example, by adding a UUID into the structure.

\section{Conclusion}
\label{s:conclusion}

In this paper we presented, Zest, a middleware solution used in the Databox project. A key design goal of Zest was to abstract complexities that exist implementing the CoAP protocol by building on top of ZeroMQ to support both a traditional REST interface together with additional high-level features such as brokering asynchronous communication between nodes. Access to all resources takes place across paths which can be controlled through a Macaroon minting process. This level of access control is valuable to IoT deployments such as Databox. The work presented in this paper is open-source and available for download at \url{https://github.com/me-box/zestdb} under an MIT license.

\section{Acknowledgements}

Work funded by EPSRC grants EP/N028260/1, EP/M001636/1 and EP/M02315X/1. Andr\'es Arcia-Moret is collaborating with Databox project under the support of Grant  RG90413 NRAG/527.

\bibliographystyle{IEEEtran}
\bibliography{zestnew}

\end{document}